# Semiconductor-metal phase transition and emergent charge density waves in 1$T$-ZrX$_2$ (X = Se, Te) at the two-dimensional limit

Ming-Qiang Ren[1], Sha Han[1], Jia-Qi Fan[1], Li Wang[2], Pengdong Wang[2], Wei Ren[2], Kun Peng[2], Shujing Li[3], Shu-Ze Wang[1], Fa-Wei Zheng[4], Ping Zhang[3], Fangsen Li[2,*], Xucun Ma[1,5,*], Qi-Kun Xue[1,5,6,*], Can-Li Song[1,5*]

[1]State Key Laboratory of Low-Dimensional Quantum Physics, Department of Physics, Tsinghua University, Beijing 100084, China

[2]Vacuum Interconnected Nanotech Workstation, Suzhou Institute of Nano-Tech and Nano-Bionics, Chinese Academy of Sciences, Suzhou 215123, China

[3]College of Mathematics and Physics, Beijing University of Chemical Technology, Beijing 100029, China

[4]Institute of Applied Physics and Computational Mathematics, Beijing 100088, China

[5]Frontier Science Center for Quantum Information, Beijing 100084, China

[6]Beijing Academy of Quantum Information Sciences, Beijing 100193, China

Corresponding*:clsong07@mail.tsinghua.edu.cn, fsli2015@sinano.ac.cn, xucunma@mail.tsinghua.edu.cn, qkxue@mail.tsinghua.edu.cn

**Abstract:** Charge density wave (CDW) is a collective quantum phenomenon in metals and features a wave-like modulation of the conduction electron density. A microscopic understanding and experimental control of this many-body electronic state in atomically thin materials remain hot topics in materials physics. By means of material engineering, we realized a dimensionality and Zr intercalation induced semiconductor-metal phase transition in $1T$-ZrX$_2$ (X = Se, Te) ultra-thin films, accompanied by a commensurate $2 \times 2$ CDW order. Furthermore, we observed a CDW energy gap up to 22 meV around the Fermi level. Fourier-transformed scanning tunneling microscopy and angle-resolved photoemission spectroscopy reveal that $1T$-ZrX$_2$ films exhibit the simplest Fermi surface among the known CDW materials in TMDCs, consisting only of Zr 4d-derived elliptical electron conduction band at the corners of the Brillouin zone.

**Keywords:** Charge density wave, semiconductor-metal transition, interface engineering, thin films, TMDC

Layered transition metal dichalcogenides (TMDCs) constitute a promising platform for exploring the effect of dimensionality on correlated electronic phases such as charge density wave (CDW) order.[1-12] Distinct from many quasi-one-dimensional systems (e.g. $ZrTe_3$, $NbSe_3$, and organic salts), in which nesting between parallel sections of the Fermi surface (FS) drives the CDW order *via* a Peierls instability, the CDW is more complex in the two-dimensional (2D) TMDCs.[13-16] Different patterns or wave vectors ($q_{CDW}$) coexist even within the same family of TMDCs.[17,18] An archetypal example of $1T$-$TiSe_2$ ($TiSe_2$ in short) undergoes the CDW phase transition with a commensurate $2 \times 2 \times 2$ superstructure below $T_{CDW} \sim 205$ K, followed by a partial gap opening at the Fermi level ($E_F$) and a chiral CDW.[19-21] By Cu intercalation or pressure application, the CDW order is suppressed and a dome-shaped superconductivity develops, while CDW patches or incommensurate CDW order could even coexist with the superconductivity.[22-25] Recently, the CDW correlations have been found to enhance in single layer $TiSe_2$ and to emerge interestingly in the closely related compound $1T$-$TiTe_2$ at the 2D limit.[26-28]

Although much has been learned about the CDWs in low-dimensional solids, a coherent understanding and control of the CDW state has not yet emerged.[4,29,30] In particular, a consistent picture is still to develop over how the CDW state evolves with reducing material thickness down to the two-dimensional (2D) limit. In $TiSe_2$, the Fermi surface consists of a Se $4p$-derived valence band maximum (VBM) at the $\Gamma$ point and a Ti $3d$-derived conduction band minimum (CBM) at the M points in the unfolded Brillouin zone (BZ).[31-39] The CDW in $TiSe_2$ is generally believed to be closely related to the interaction between the two bands. Here, we report a more intriguing CDW state in two sister compounds, octahedral $1T$-$ZrX_2$ (X = Se, Te) films, epitaxially prepared on graphitized SiC(0001) substrates (Figure 1a) by molecular beam epitaxy (MBE). Distinct from TiSe$_2$, the pristine ZrSe$_2$ exhibits a semiconductor band gap up to 1.1 eV and is thus not expected to harbor any CDW phase.[31,34,40,41] However, it comes as a large surprise that our study by scanning tunneling microscopy and spectroscopy (STM/STS) reveals the emergence of a commensurate (2 × 2) CDW order following a semiconductor-metal transitions (SMT) in ultra-thin ZrX$_2$ (X = Se, Te) films (Figure 1b). Angle resolved photoemission spectroscopy (ARPES) measurements and fast Fourier transform (FFT) analysis of the tunneling conductance dI/dV maps reveal that the ZrX$_2$ exhibits a much simplified FS topology with only one band (Zr 4$d$-derived band) across the $E_F$, forming six elliptical electron pockets at the 2D Brillouin-zone corner (M) (see the inset of Figure 1c). The CDW and SMT transition is experimentally realized by engineering the charge transfer between the ZrX$_2$ and graphene heterointerface as well as Zr intercalation. Our findings raise several salient questions of the CDW mechanisms in general, and reveal a feasible way by interface engineering to realize the CDW states in 2D semiconductor TMDCs.

Layered 1$T$-ZrX$_2$ (X = Se, Te) crystals are made from the stacking X-Zr-X sandwiches, every of which could be viewed as a unique triple layer (TL). In our experiments, 1$T$-ZrX$_2$ (X = Se, Te) films with thicknesses ranging from 1 TL to 10 TL have been prepared and explored. Figure 1b shows a representative STM topography acquired on a 2 TL ZrSe$_2$ epitaxial film, presenting a Se-terminated hexagonal lattice decorated with sparsely distributed native defects (Figure S1), in analogy with TiSe$_2$.[4,42-44] The in-plane lattice parameter is measured to be 3.79 ± 0.10 Å (3.77 ± 0.10 Å) in 1 TL (2 TL) ZrSe$_2$, in good agreement with its bulk value of 3.77 Å.[45] This implies a negligible effect of the epitaxial strain, imposed by the underlying graphene substrate *via* a weak van der Waals interaction, on the ZrSe$_2$ films explored.

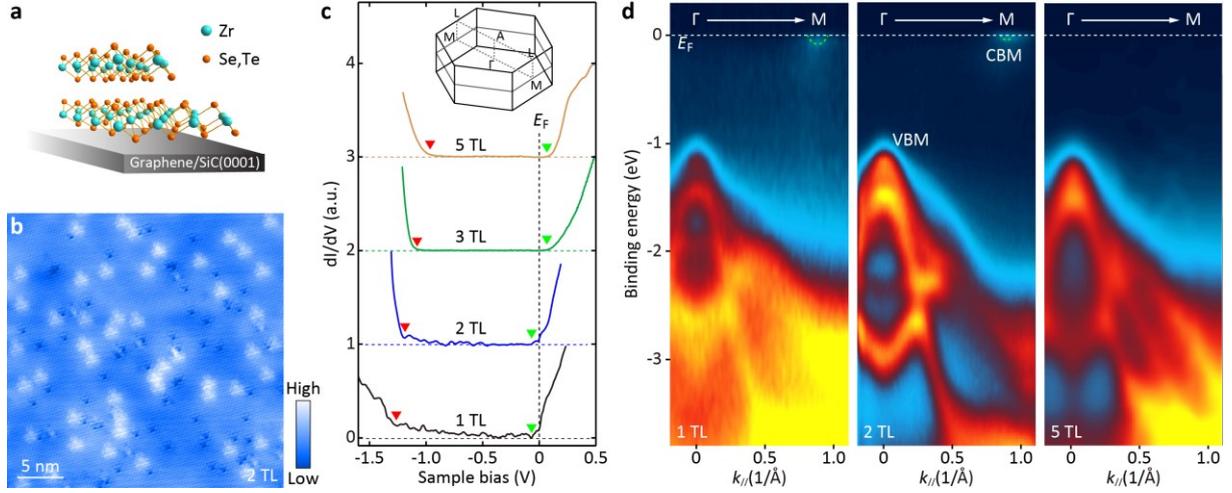

**Figure 1.** Dimensionality controlled SMT in 1$T$-ZrSe$_2$. (a) Sketch of ZrX$_2$ (X = Se, Te) films on the graphene/SiC(0001) substrate. (b) STM topography (32 nm × 32 nm, $V$ = 0.5 V, $I$ = 50 pA) on 2 TL ZrSe$_2$. (c) Spatially averaged conductance dI/dV spectra on 1 TL (black), 2 TL (blue) and 3 TL (green) and 5 TL (brown) ZrSe$_2$, measured at locations away from defects. Set point: $V$ = 0.5 V, $I$ = 100 pA, $\Delta V_{0\text{-peak}}$ = 10 mV. The green and red triangles mark the energy positions of CBM and VBM that are identified by the DOS kinks in the dI/dV spectra, while the gray dashes denote $E_F$ throughout. The finite DOS associated with MIGS emerge within the semiconductor band gaps of 1 TL and 2 TL ZrSe$_2$. Inset shows the three-dimensional Brillouin zone of ZrX$_2$ (X = Se, Te). (d) ARPES spectra of ZrSe$_2$ films at varying thickness, measured along the Γ-M direction, revealing a SMT down to the 2D limit. The green parabolas denote the conduction bands near $E_F$.

The spatially averaged conductance dI/dV spectra, which probe the quasiparticle density of states (DOS), acquired on the defect-free regions of ZrSe$_2$ films, are plotted in Figure 1c. One immediately sees a semiconductor band gap of approximately 1.2 eV. By identifying the VBM at the Γ point and CBM at the M points in the unfolded Brillouin zone (BZ) as the DOS kinks in the dI/dV spectra, marked by the green and red triangles, respectively, we reveal a downward shift of CBM as the ZrSe$_2$ becomes thinner. Although the CBM slightly alters in energy with varying spatial positions, it has always sunken below $E_F$ for 1 TL and 2 TL ZrSe$_2$. This leads to a finite DOS around $E_F$ and a transition from semiconductor to metal phase down to the 2D limit. Such an electronic phase transition is convincingly confirmed by thickness-dependent ARPES spectra along the $\Gamma$ - M direction in Figure 1d. The conduction band, which is missing at $E_F$ in 5 TL ZrSe$_2$, emerges at M points and crosses $E_F$ in both 1 TL and 2 TL ZrSe$_2$ films, as sketched by the green dashed parabolas. This contributes a tiny electron pocket at the six-shared M points of the Brillouin zone. According to previous ARPES measurements and theoretical calculations, the conduction (M) and valence ($\Gamma$) bands are derived from Zr 4$d$ and Se 4$p$ orbitals, respectively.[40,41,45-49] The indirect $p$-$d$ bandgap extracted from our ARPES intensity is about 1.1 eV, largely consistent with the STS measurements in Figure 1c. For 1 TL and 2 TL ZrSe$_2$ films, the binding energy of VBM shows a small discrepancy between the STS (4.5 K) and ARPES (300 K) measurements, which is most probably caused by the temperature difference used. At low temperature, the emergent CDW order might push the valence states to lower energies, thereby reducing the total electronic energy of system.[31,37] In addition, finite DOS is found in the semiconducting bandgap of 1 TL and 2 TL ZrSe$_2$ (Figure 1c), which may be related to metal-induced in-gap states (MIGS, see Figure S1) in van der Waals metal-semiconductor junction.[50]

The most likely explanation for the observed SMT is the charge transfer at the ZrSe$_2$/graphene interface. This makes sense since the ZrSe$_2$ has a larger work function of $\sim$ 5.2 eV than that of graphene (4.2 - 4.4 eV).[51] Upon contact, electrons flow through the junction from graphene to ZrSe$_2$ and the accumulated electrons would lift $E_F$ of the system across the CBM, prompting the SMT at the 2D limit. We also note that there three types of native defects (Se$_{up}$ vacancies, intercalated Zr atoms and Se$_{down}$ defects) in ZrSe$_2$ films, which are all electron donors (Figure S1). However, the intrinsic doping effect of the defects is not decisive to the SMT in ZrSe$_2$ as noted in Figure S1.

Following the SMT, we observe the emergence of a commensurate (2 × 2) CDW order on both 1 TL and 2 TL ZrSe$_2$, as demonstrated in Figures 2a and 2b. In analogy to TiSe$_2$, three neighboring Se atoms distort into nearly-circular patterns in one 2 × 2 supercell (marked by the white rhombuses), while the fourth Se atom is relatively darker and essentially unchanged.[43] The CDW modulation appears more prominent in 2 TL ZrSe$_2$ compared to that in 1 TL films, where CDW nucleates locally around the defects. The spatial 2 × 2 modulation is discernible as six bright spots in the FFT image, as marked by the white circles in Figure. 2c. Figures 2d and 2e show the zoom-in STM topographic images of a 2 TL film to reveal the expected contrast reversal, which serves as a signature of CDW.[52-54] Accompanied with the CDW modulation are partial energy gaps of about 22 meV that open at $E_F$ (Figure 2f). The gap magnitude is comparable to that (28 meV) of single-layer TiTe$_2$.[28] One may notice the appearance of two conductance kinks near ± 4 meV, which were also observed in TiTe$_2$, but their origin remains unknown to date.[28]

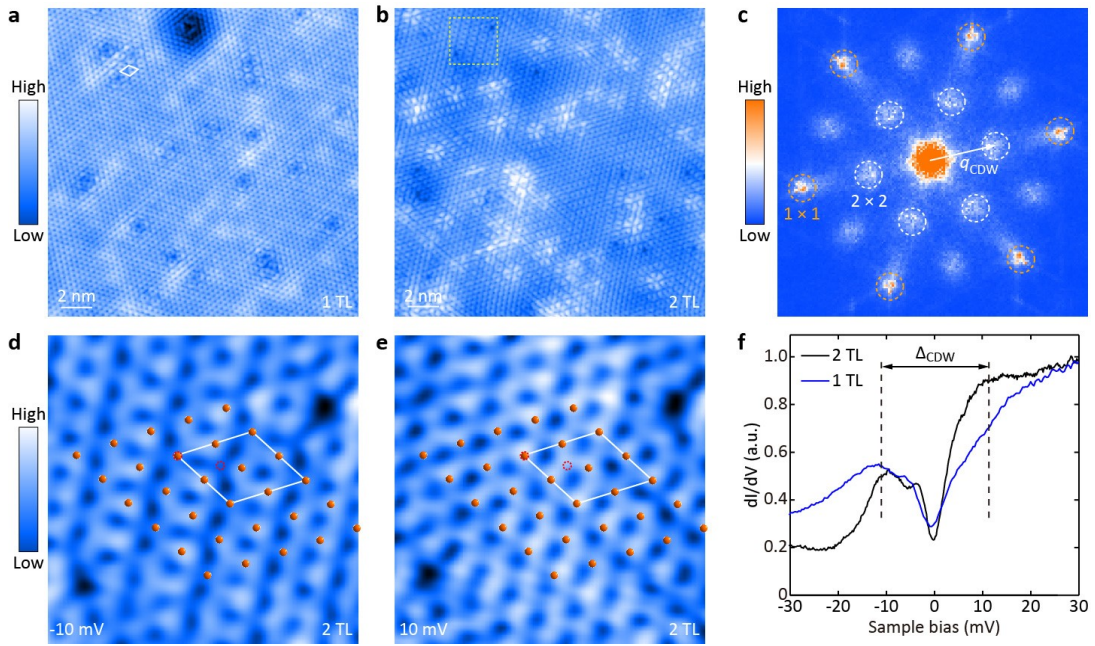

**Figure 2.** Emergent CDW in 1 TL and 2 TL ZrSe$_2$. (a,b) STM topographies (20 nm × 20 nm, $V$ = -10 mV, $I$ = 50 pA) showing a 2 × 2 CDW order in 1 TL and 2 TL ZrSe$_2$ films, with the white rhombuses marking the supercell of CDW modulation. (c) FFT image of (b). The 1 × 1 Bragg peaks and 2 × 2 CDW spots are circled in orange and white, respectively. (d,e) Filled- and empty-state STM images of 2 TL ZrSe$_2$ film (3 nm × 3 nm, $V$ = ±10 mV, $I$ = 50 pA) taken within the yellow box in (b), showing a contrast inversion between opposite bias polarity (e.g. the red circle-marked sites). The orange spheres denote the undistorted Se lattice Se at the top surface. (f) Spatially averaged dI/dV spectra ($V$ = 30 mV, $I$ = 200 pA, $\Delta V_{0\text{-peak}}$ = 1 mV) exhibiting CDW gaps in 1 TL (blue) and 2 TL ZrSe$_2$ (black).

To understand the underlying origin of the CDW, we investigate the electronic structure around $E_F$ in 1T-ZrSe$_2$. As discussed above, the Zr 4$d$ conduction band of ZrSe$_2$ is tuned to across $E_F$ with reduced dimensionality, while the Se 4$p$ valence bands are located far below $E_F$ (Figures 1c and 1d). This results in only electron pockets at the M points of the 2D BZ around $E_F$ in 1 TL and 2 TL ZrSe$_2$. We corroborate such a FS contour by Fourier-transforming the spectroscopic dI/dV maps in Figures 3a-j, measured on one 1 TL ZrSe$_2$ film. The FFT images contain three sets of elliptic scattering rings as sketched by $q_1$ (green), $q_2$ (orange) and $q_3$ (red), respectively. Evidently, the three scattering rings are dispersive and enlarge with increasing energy. Figure 3i depicts the FFT intensity along the Γ - M direction, clearly revealing electron-like dispersions of $q_1$ and $q_3$. QPI patterns originate from scatterings between electronic states on a constant energy contour in the momentum ($k$) space. By considering the simple FS topology in Figures 3j and 3k and calculating the noninteracting joint density of states $\text{JDOS}(q) = \int I(k)I(k+q)\text{d}^2k$, we simulate the FFT image in Figure 3l.[55] One sees that all three sets of scattering rings in Figures 3e-h are mostly reproduced in the simulated FFT image, including their orientations and locations in the reciprocal space. A minor discrepancy is that the intensity of the scattering rings in Figures 3e-h display some inhomogeneity, probably due to the matrix effect or the DOS hotspots that are neglected in the simulation. Clearly, the three sets of scattering rings stem from the nearest neighboring ($q_1$), next nearest neighboring ($q_2$) and the third nearest neighboring ($q_3$) intra-pocket scatterings between the M electron pockets (conduction band), respectively, as sketched in Figure 3k. To best reproduce the FFT images, an anisotropic ratio ($k_{\Gamma\text{-M}}/k_{\text{M-K}}$) of the elliptical electron pockets is determined to be 2.3:1, a value consistent with the ARPES measurements and theoretical calculations.[40,41,47,48] However, the nearly rounded pocket in our ARPES may arise from the very tiny pocket size and matrix effect in our films. The energy position of CBM for 1 TL ZrSe$_2$ is extracted to be about 30 meV underneath $E_F$ from the FFT intensity cut (Figure 3i). This agrees with the dI/dV spectra in Figure 1c, in which the DOS is profoundly increased at an almost same energy. We also note that features of $q_1$ and $q_{\text{CDW}}$ merge together around the M points in the FFT images as the energy approaches CBM (Figures 3a and 3e). Further experiments are needed to disentangle between $q_{\text{CDW}}$ and $q_1$. Nevertheless, the emergence of CDW is confirmed in thin ZrSe$_2$ films with a rather simple FS, consisting of only an electron pocket around M.

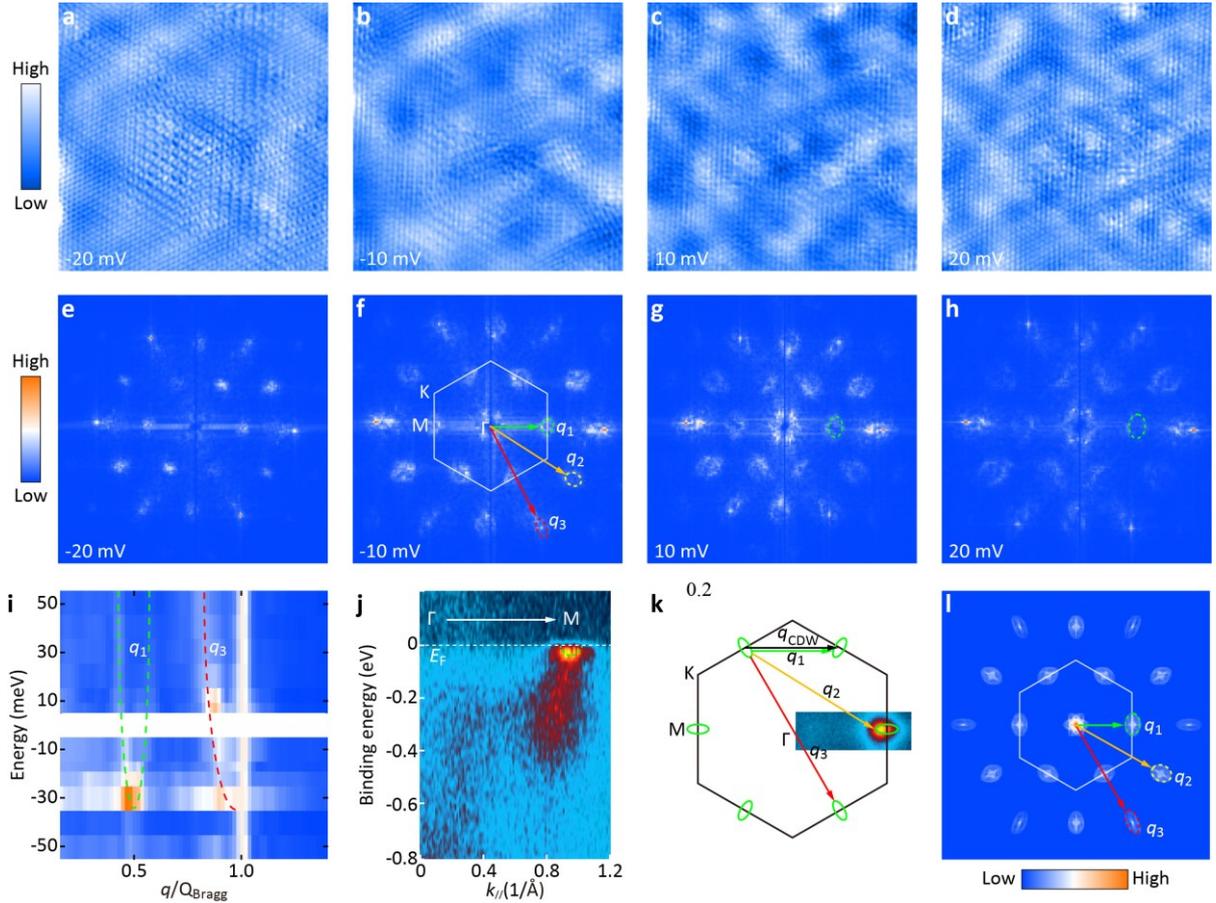

**Figure 3.** FS topology. (a-d) Spectroscopic dI/dV maps taken on an 18 × 18 nm area of 1 TL ZrSe₂ with various biases (as labelled). (e-f) FFT images transformed from (a-d), respectively. The white hexagons sketch out the unfolded 2D BZ of ZrSe₂. The wavevectors of $q_1$ (green) and $q_2$ (yellow) and $q_3$ (red) mark three distinctive sets of scattering rings. (i) FFT line cuts extracted along the Γ-M direction, which is normalized by the intensity at Bragg peaks. The dispersive scattering rings ($q_1$ and $q_3$) are sketched out with the green and red dashed curves. (j,k) Zoom-in ARPES spectrum ($T$ = 30 K) along the Γ-M direction and FS map acquired on 1 TL ZrSe₂. The FS map is obtained by integrating the spectral intensity within ± 10 meV of $E_F$. Schematic diagram of the 2D BZ and the electron pockets at M are shown for clarify in (k). The $q_1$, $q_2$, $q_3$ and $q_{CDW}$ label the distinctive intra-pocket scattering vectors between the M electron pockets and the CDW wave vector, respectively. (l) Calculated joint DOS at 10 meV by considering the FS in (k), with only electron pockets at M.

To confirm the compatibility of the commensurate $2 \times 2$ CDW with a simple Fermi surface, we further explored ultra-thin 1$T$-ZrTe$_2$ films. Distinct from the semiconductor ZrSe$_2$, the sister compound ZrTe$_2$ was mostly discussed as a semimetal.[41,45,56,57] However, our STM and STS measurements of ZrTe$_2$ thin films reveal a semiconductor-like band structure, as illustrated in Figures 4a-d. This is essentially true for 1 TL ZrTe$_2$, in which an obvious band gap of approximately 0.37 eV is observable (top panel in Figure 4d), irrespective of the native defects (Figure S2). An FFT of the dI/dV mapping of 1 TL ZrTe$_2$ measured at zero bias also gives rise to elliptical scattering rings in Figure 4e, bearing close resemblance to those of 1 TL ZrSe$_2$ (Figures 3e-f). By employing a similar FS with only electron pockets at the M points and setting a larger anisotropy $k_{\Gamma\text{-M}}/k_{\text{M-K}}$ of 3.1 (Figure 4f), the calculated JDOS nicely reproduces the experimental scattering features as well (Figure 4g). In brief, our results show that 1 TL ZrTe$_2$ has a similar simple FS topology as $ZrSe_2$, due to the interface-induced SMT. A recent ARPES measurement of ultra-thin $ZrTe_2$ films grown on InAs(111) reveals a Dirac-like cone at $\Gamma$, hinting that $ZrTe_2$ could be a potential topological Dirac semimetal candidate.[56,57] Such an intriguing electronic state seems absent in 1 TL $ZrTe_2$ films on graphene. In contrast to the massless Dirac fermions, our STS reveals sharply increasing DOS below -0.4 eV (Figure 4d), matching the previously identified parabolic valence band at - 0.4 ~ -0.5 eV by ARPES.[56,57]

The situation becomes a little complicated in thicker $ZrTe_2$ films. Due to a larger van der Waals gap in $ZrTe_2$, a considerable number of excess Zr atoms ($Zr_{inter}$) intercalate between the sheets of $ZrTe_2$ and mess up the surface so badly (Figures 4b and 4c).[57] At the same time, the dI/dV spectra are somewhat vague, albeit with the gap-like feature, and an unambiguous determination of their semiconductor or semimetal nature seems impossible. Fortunately, the peak-like and hump-like dI/dV structures from the Zr 4$d$ conduction and Te 4$p$ valence bands, denoted by the green and red arrows in Figure 4d, respectively, could be readily distinguished and are nearly evenly spaced (~ 1.2 eV). This means a robust band structure of $ZrTe_2$ that alters little with the film thickness. In this context, the downward shift of the energy bands for both 2 TL and 7 TL with respect 1 TL matches a heavy electron doping by the considerable $Zr_{inter}$ defects there. We are therefore convinced that the $E_F$ of 2 TL and 7 TL $ZrTe_2$ films has been lifted above the CBM and only cuts through the Zr 4d conduction band, just like ultrathin $ZrSe_2$ films. Note that the occurrence of the heaviest electron doping in 2 TL $ZrTe_2$ might be a combined effect of the $Zr_{inter}$ defects and interfacial electron transfer.

Having identified a similar FS topology, we examine the possibility of CDW in $ZrTe_2$. Unlike 1TL $ZrSe_2$, 1 TL $ZrTe_2$ shows no short-range CDW in the vicinity of native defects, possibly owing to the weak electron donor capacity (Figure S2). In some region, unidirectional CDW stripe is detectable (Figure S3), whose origin merits further study. Such a stripe-like CDW has previously been observed in the NbSe$_2$ and Cu-intercalated TiSe$_2$.[58,59] Overall, the CDW turns out to be comparably weak in 1 TL ZrTe$_2$. Intriguingly, as the apex of STM tip is functionalized possibly by attaching a Te atom from the ZrTe$_2$ surface *via* applying a positive sample pulse, it becomes less sensitive to the Zr$_{inter}$ defects and a $2 \times 2$ CDW order can be repeatedly observed in thicker ZrTe$_2$ films, as exemplified for a 7 TL film in Figure 4h. Figure 4i zooms into the associated lattice distortion in the CDW phase, where the shrinkage of three adjacent Te atoms (marked by the white arrows) in one $2 \times 2$ supercell is more clearly evident. All the observations support an identical CDW order in 1$T$-ZrX$_2$ (X = Se, Te), albeit a smaller CDW energy gap of $\sim 11$ meV in ZrTe$_2$ (Figure 4j).

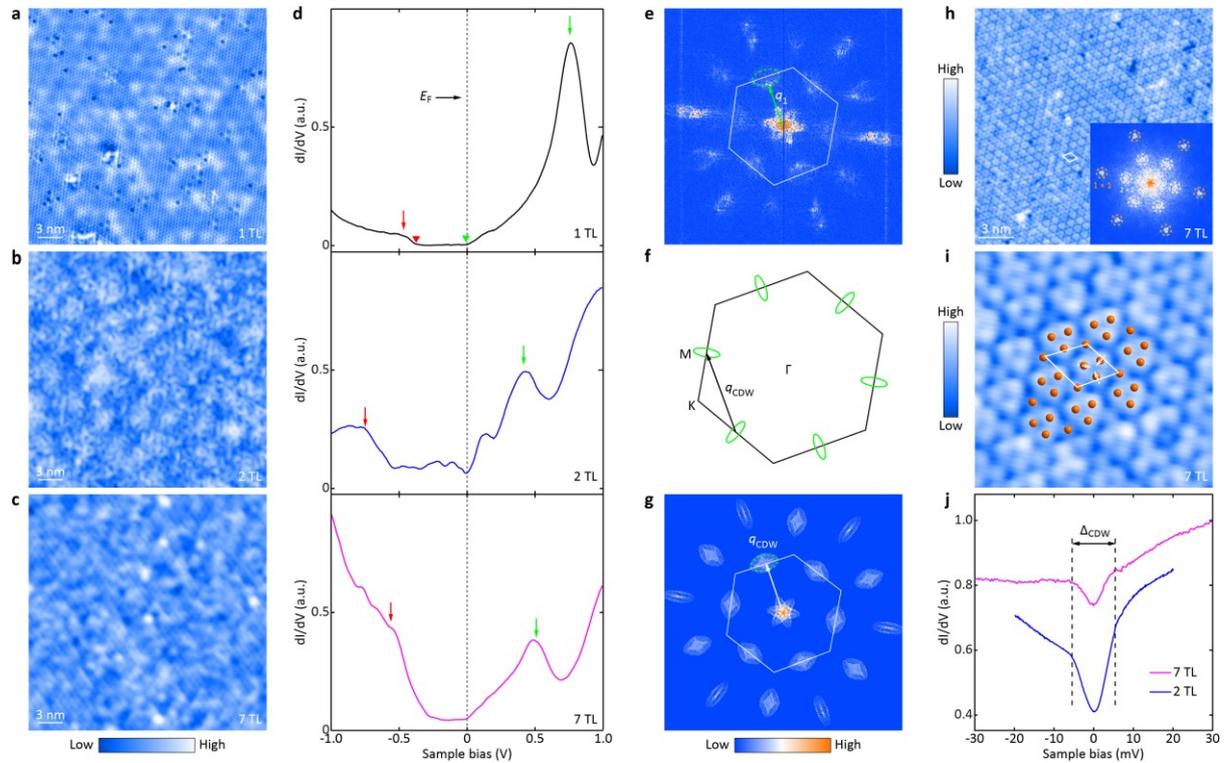

**Figure 4.** CDW order in 1$T$-ZrTe$_2$. (a-c) STM topographies (25 nm × 25 nm, $V = 0.1$ V, $I = 50$ pA) on 1 TL, 2 TL and 7 TL ZrTe$_2$ films, decorated by sparse native defects. (d) Spatially

averaged dI/dV spectra ($V = 1.0$ V, $I = 200$ pA, $\Delta V_{0\text{-peak}} = 30$ mV) on 1 TL (top), 2 TL (middle) and 7 TL ZrTe$_2$ (bottom). The green and red arrows represent the electronic DOS from the Zr 4d conduction and Te 4p valence bands, respectively. (e) FFT image transformed from a zero-bias STS mapping taken on a $35 \times 35$ nm area of 1 TL ZrTe$_2$. (f,g) FS topology and calculated JDOS with a larger anisotropy $k_{\Gamma\text{-M}}/k_{\text{M-K}}$ of 3.1 of the elliptical pockets. (h) STM topography (20 nm × 20 nm, $V = 20$ mV, $I = 100$ pA) displaying a similar 2 × 2 CDW modulation on 7 TL ZrTe$_2$ by a functionalized STM tip, with the inset showing its FFT image. The FFT peaks of CDW modulation are white-circled at $q_{\text{CDW}} = Q_{\text{Bragg}}/2$. (i) Magnified view of the CDW-associated lattice distortion (4 nm × 4 nm, $V = 20$ mV, $I = 100$ pA). The orange spheres denote the top Te atoms. (j) Small-energy-scale STS showing a CDW energy gap of about 11 meV on 2 TL and 7 TL ZrSe$_2$. Setpoint: $V = 30$ mV, $I = 100$ pA, $\Delta V_{0\text{-peak}} = 1$ mV.

Although the emergent (2 × 2) CDW order and lattice distortion pattern are reminiscent of those of TiSe$_2$,[4,17,19-23,27-39,43] their revelation in the two ZrX$_2$ (X = Se, Te) films with a simple FS topology is much more intriguing. First, the CDW mechanisms invoking the electron-hole band interaction, such as the excitonic insulator scenario and Jahn-Teller band instability seems unlikely for the CDW formation in ZrX$_2$, as their hole pockets at $\Gamma$ are completely missing around $E_{\text{F}}$ in ZrX$_2$ (X = Se, Te) (Figures 3 and 4e-g).[28,31,32,33-37,60] This argument was supported by a recent report on TiSe$_2$, in which the CDW-associated periodic lattice distortion and energy gap survive even when the excitonic order is quenched.[39] Second, the conduction and valence bands are well separated by a semiconductor band gap of ~ 1.2 eV in ZrSe$_2$ (Figure 1c). The traditional Fermi surface nesting scenario between the two $\Gamma$- and M-centered bands is thus inapplicable for the CDW formation in ZrX$_2$ as well. Instead, we note that the scattering vectors between the consecutive electron pockets are close to the $q_{\text{CDW}}$ of the CDW modulation (Figures 3k and 4f), especially for a tiny FS on moderate doping. This renders it possible to ascribe the CDW formation to nested FS pockets at the M points. Such a novel CDW mechanism is consistent with the opened energy gaps at $E_{\text{F}}$ (Figures 2f and 4j) and emergent

CDW in the Cu-intercalated TiSe$_2$, where the Se 4p hole pockets are pushed far below $E_F$ and only inter-valley scatterings between the electron pockets at M are possible.[61] Last but more importantly, the present study realizes a rare observation of CDW in semiconductor transition metal dichalcogenide (TMDC), and reveals that the present Zr 4$d$ band itself at $E_F$ is a necessary and sufficient condition for the CDW formation in 1$T$-ZrX$_2$, although their bulk counterparts are semiconductors and exhibit no CDW.[41]

Looking ahead, a couple of interesting questions merit further investigations. The first and most significant one is why the CDW correlations are so strong around native defects in 1 TL ZrSe$_2$ films and occur only locally in 1 TL ZrTe$_2$ films. The defect-mediated local doping and therefore CDW condensation might be the possible causes for it as well as the robust CDW in 2 TL ZrSe$_2$ and multilayer ZrTe$_2$ films.[41] This comes to another prominent concern how the CDW of ZrX$_2$ (X = Se, Te) evolves with doping level and interplays with other collective phenomena such as superconductivity observed in alkali-metal intercalated ZrSe$_2$ and predicted in Ni-doped ZrTe$_2$.[64,65] Anyhow, our results reveal a unique compatibility between the simple FS topology and the CDW phase, and open new avenues towards creating and controlling the many-body CDW phase in semiconductors. Apparently, dimensionality and interface engineering of TMDCs, including the semiconducting ones, may discover more exotic CDW systems and shed important light on the CDW mechanism in the layered materials.

**Methods**

**Sample preparations** Our experiments were conducted in two commercial ultra-high vacuum STM facilities (Unisoku 1300 and 1500), integrated with two separate MBE chambers for *in-situ* film preparation. The base pressure of all chambers is lower than $2.0 \times 10^{-10}$ Torr. Ultrathin $ZrX_2$ (X = Se, Te) films were epitaxially grown on nitrogen-doped 6H-SiC(0001) wafers with a resistivity of < 0.1 $\Omega \cdot$cm (TankeBlue Semiconductor Co. Ltd.), which were pre-graphitized by heating to 1350$^{\circ}$C for 10 minutes. This graphitization method often leads to the formation of dominant double-layer graphene on the surface.[66] High purity Zr (99.999%) and (Se/Te) (99.999%) were co-evaporated from a single-pocket electron beam evaporator (SPECS) and standard Knudsen diffusion cells, respectively. The typical growth temperatures for both $ZrSe_2$ and $ZrTe_2$ films are 700$^{\circ}$C, with a high X/Zr (X = Se, Te) flux ratio of > 10. The self-regulating stoichiometry control for $ZrX_2$ (X = Se, Te) films closely resembles the MBE growth dynamics of other metal chalcogenides such as $TiSe_2$.[4]

**STM measurements** After the MBE growth, as-prepared $ZrX_2$ (X = Se, Te) epitaxial films were immediately transferred into the STM modules for all STM and STS data collections at 4.5 K. A bias voltage was applied to the samples. Polycrystalline PtIr tips were conditioned by e-beam bombardment in MBE, calibrated in Ag/Si(111) films and used throughout the experiments. All STM topographies were acquired in a constant current mode. The tunneling spectroscopy (dI/dV) and conductance map are measured using a standard lock-in technique that applied an additional small a.c. voltage with a peak value $\Delta V$ and a frequency $f$ = 983 Hz.

**ARPES measurements** Our ARPES measurements were performed with a photon energy of 21.2 eV at room temperature, unless otherwise specified. Before being transferred into ARPES chamber, the as-prepared $ZrSe_2$ films were coated by several nanometers Se layers. Then, we annealed the $ZrSe_2$ films at 300$^{\circ}$C to remove the Se protection layers. This method has proved feasible to maintain the atomic and electronic structures of $ZrSe_2$ films by STM, as noted in Figure. S4.

ASSOCIATED CONTENT

**Supporting Information:** The Supporting Information is available free of charge online.

Native defects, formation mechanism and their effects on SMT; Native Te defects in monolayer ZrTe$_2$; Unidirectional CDW modulation in 1 TL ZrTe$_2$; Decapping Se protection layers for *ex-situ* ARPES measurements; etc.

AUTHOR INFORMATION

**Author contributions:** C.L.S., X.C.M. and Q.K.X. conceived and designed the experiments. M.Q.R. and S.H. carried out the MBE growth and STM measurements with assistance from J.Q.F., L.W., W.R., K.P., S.Z.W., P.D.W. and F.S.L. performed the APRES measurements. M.Q.R. and C.L.S. analyzed the data and wrote the manuscript with comments from S.J.L., F.W.Z., P.Z., X.C.M., Q.K.X.

**Notes:**

The authors declare no competing financial interest.

ACKNOWLEDGMENT

The work was financially supported by the Ministry of Science and Technology of China (2017YFA0304600, 2016YFA0301004, 2018YFA0305603), the Natural Science Foundation of China (Grants No. 51788104, No. 11634007, and No. 11774192), Nano-X from Suzhou Institute of Nano-Tech and Nano-Bionics (SINANO), Chinese Academy of Sciences, and in part by the Beijing Advanced Innovation Center for Future Chip (ICFC).

# Semiconductor-metal phase transition and emergent charge density waves in 1*T*-ZrX$_2$ (X = Se, Te) at the two-dimensional limit

Ming-Qiang Ren[1], Sha Han[1], Jia-Qi Fan[1], Li Wang[2], Pengdong Wang[2], Wei Ren[2], Kun Peng[2], Shujing Li[3], Shu-Ze Wang[1], Fa-Wei Zheng[4], Ping Zhang[3], Fangsen Li[2*], Xucun Ma[1,5,*], Qi-Kun Xue[1,5,6*], Can-Li Song[1,5*]

[1]State Key Laboratory of Low-Dimensional Quantum Physics, Department of Physics, Tsinghua University, Beijing 100084, China

[2]Vacuum Interconnected Nanotech Workstation, Suzhou Institute of Nano-Tech and Nano-Bionics, Chinese Academy of Sciences, Suzhou 215123, China

[3]College of Mathematics and Physics, Beijing University of Chemical Technology, Beijing 100029, China

[4]Institute of Applied Physics and Computational Mathematics, Beijing 100088, China

[5]Frontier Science Center for Quantum Information, Beijing 100084, China

[6]Beijing Academy of Quantum Information Sciences, Beijing 100193, China

Corresponding*:clsong07@mail.tsinghua.edu.cn, fsli2015@sinano.ac.cn, xucunma@mail.tsinghua.edu.cn, qkxue@mail.tsinghua.edu.cn

## Table of Contents

**Note S1. Native defects, formation mechanism and their effects on SMT**

Figures 1a,b present the atomically resolved STM topographies of 1 TL 2 TL ZrSe$_2$ films, respectively. Based on the morphologies, one immediately notes three different kinds of defects, as circled by the white dashes. These defects have been commonly revealed in both 1$T$-TiSe$_2$ crystals and thin films and their origins are also investigated in detail.[1-3] Type I and type III defects are assigned to Se$_{up}$ vacancies and intercalated transition metal atoms inside the van der Waals gap of adjacent layers. Type II defects occupy at the down-layer Se sites and are ascribed as Se$_{down}$ substitutions by residual iodine and oxygen or Se interstitials.[1-3] Nevertheless, all the three type of defects are electron donors as demonstrated by the STS measurements (Figure S1c). The characteristic peak (marked by the light blue arrows) derived from the Se 4$p$ states uniformly moves to a lower energy on single defects relative to the defects-free regions. Despite the donor nature, the native defects fail to account for the SMT identified in ultrathin ZrSe$_2$ films, because the contradiction that monolayer ZrSe$_2$ is the most conductive (Figures 1c and 1d) but processes least defects (Figure S1d). We note that the Zr$_{inter}$ induces a defect state around -0.5 eV, possibly related to the non-dispersive spectral weight in Cu intercalated ZrSe$_2$.[4]

At the metal-semiconductor junction, the MIGSs are caused by the decay of complex wave function from the metal into the band gap of the semiconductor and decay exponentially away from the interface. According to the one-dimensional Monch model, the minimum MIGS decay length could be easily estimated as $\delta_{min} = \frac{\hbar^2}{m_e a E_g} \sim 0.70$ nm, in which $\hbar$ is the Planck constant, $m_e$ is the free electron mass, $a$ = 3.77 Å and $E_g$ = 1.2 eV are the lattice parameter and semiconductor band gap of ZrSe$_2$, respectively.[5] This value is slightly greater than the out-of-plane lattice parameter of $\sim$ 0.61 nm in ZrSe$_2$, in agreement with our experimental observation of MIGSs at most in 2 TL ZrSe$_2$ (Figure. 1c). It is worth mentioning that the MIGSs serve as a reservoir for electrons or holes and usually pin $E_F$ in traditional metal-semiconductor heterojunction.[6] However, in our case, the van der waals heterostructure of ZrSe$_2$/graphene produces relatively fewer MIGSs, which, in conjunction with the sharp and high-quality interface, is not sufficient for effective pinning of $E_F$. The weak $E_F$ pinning agrees markedly with a recent theoretical calculation based on the van der Waals metal-semiconductor junction.[7]

**Note S2.  JDOS simulation**

Fourier transform scanning tunneling spectroscopy (FT-STS) technique exhibits great capability in determining the Fermi surface contour (FSC), and has been widely applied in noble metals and high temperature superconductors.[8-10] To examine the FSC of 1 TL ZrSe$_2$, we analyzed the FFT images of spectroscopic dI/dV maps taken near $E_F$ (Figures 3a-d). It is well-known that the FFT patterns originate from scatterings between the electronic states on a constant energy contour in the momentum ($k$) space, which is the joint density of states $JDOS(q) = \int I(k)I(k+q)d^2k$.[11] To calculate JDOS, we consider a simple FS topology, which consists only six elliptical electron pockets around M in Figure 3k, as detailed in the main text. Two key parameters, namely the large axes (a, namely $k_{\Gamma\text{-M}}$) and the ratio of the large and small axes (a/b, namely $k_{\Gamma\text{-M}}/k_{\text{M-K}}$) of the electron pockets, play a decisive role in the simulated patterns. During the calculation, we vary the parameters a and a/b carefully to best reproduce the experimental results, especially the outer elliptical rings as marked in Figure 3l. To obtain Figure 3l, we set a/b = 2.3:1which agrees qualitatively with the corresponding value of around 3.0 inferred from the anisotropic effective electron mass of 1TL ZrSe$_2$ by theoretical calculation.[12] Inside the elliptical ring, one might note a diamond shaped feature with strong intensity in the calculated JDOS, but disappears completely in the experimental patterns. This divergence might result from the oversimplification of our calculations, which considers an ideal electron pocket without any $k_z$ dependence, tunneling matrix effect and/or DOS hotspot.

## Note S3. Native Te defects in monolayer ZrTe$_2$

In analogy to 1 TL ZrSe$_2$ (Figure S1a), there also exist two categories of defects in 1TL ZrTe$_2$ that register at the up (Te$_{up}$) and bottom (Te$_{down}$) layer of Te atoms, shown in Figure S2a, respectively. They share similar STM morphologies with Se$_{up}$ and Se$_{down}$ defects in ZrSe$_2$ and can be reasonably assigned to native Te$_{up}$ and Te$_{down}$ defects. Distinct from 1 TL ZrSe$_2$, however, no CDW is observed in 1 TL ZrTe$_2$. A series of dI/dV spectra measured across both Te$_{up}$ and Te$_{down}$ defects (along the two purple arrows in Figure S2a) are depicted in Figures S2b and S2c. The characteristic DOS peaks at ~ 0.76 eV, CBM and VBM remain unchanged even in the presence of Te vacancies, indicative of a weak donor capacity of the native Te$_{up}$ and Te$_{down}$ defects in ZrTe$_2$.

## Note S4. Unidirectional CDW modulation in 1 TL ZrTe$_2$

Occasionally, unidirectional CDW modulation was evident in 1 TL ZrTe$_2$ films, as illustrated in the fan-shaped region of Figure S3a. The unidirectional nature of the CDW pattern has been further demonstrated by the FFT image in Figure S3b. The FFT spots of $q_{CDW}$ are observable only along one of the symmetry equivalent $\Gamma$M directions. Such a unidirectional CDW might be related to the epitaxial strain induce by certain defects (such as ripples) in the graphene substrate, which merits a future investigation.

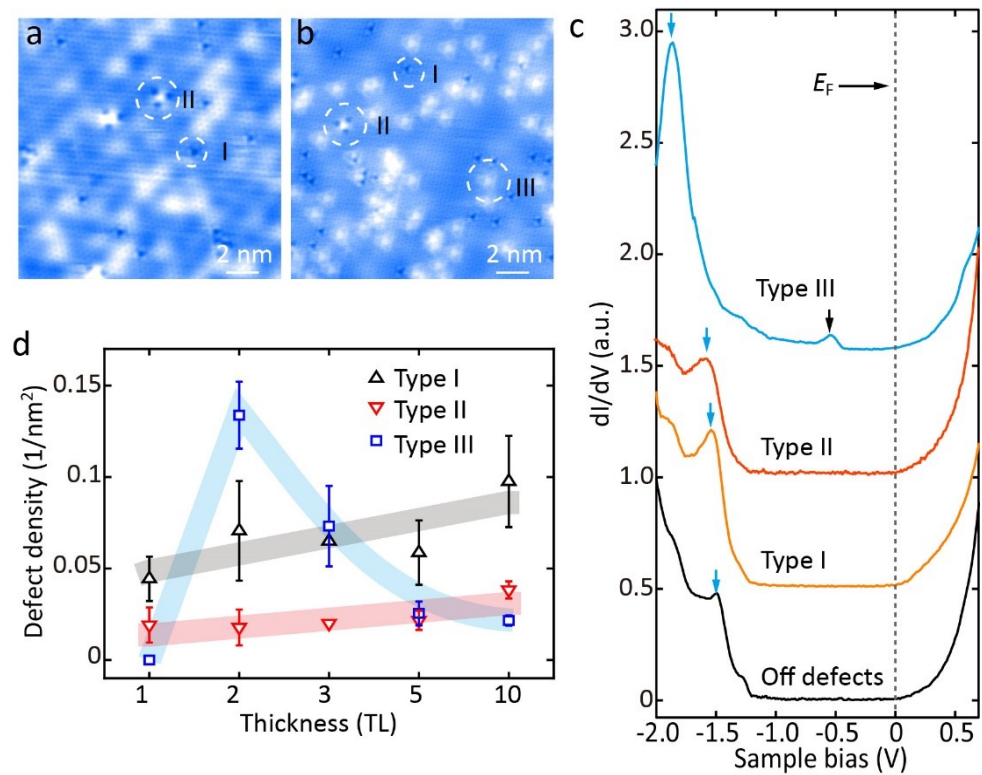

**Figure S1.** Native defects in ZrSe$_2$. (a,b) Atomically resolved STM topographic images (16 nm × 16 nm) of 1 TL and 2 TL ZrSe$_2$ films. Tunneling conditions: $V = 0.5$ V, $I = 50$ pA. The dotted circles with different sizes mark three distinct defects. (c) dI/dV spectra ($V = 1.0$ V, $I = 200$ pA, $\Delta V = 30$ mV) on single defects and regions away from defects. For comparison, all spectra were measured on one 2 TL ZrSe$_2$ film. The light blue and black arrows mark the prominent electronic states from the Se 4p valance bands and Zr$_{inter}$ defect, respectively. (d) Statistics of the areal defect densities as a function of film thickness.

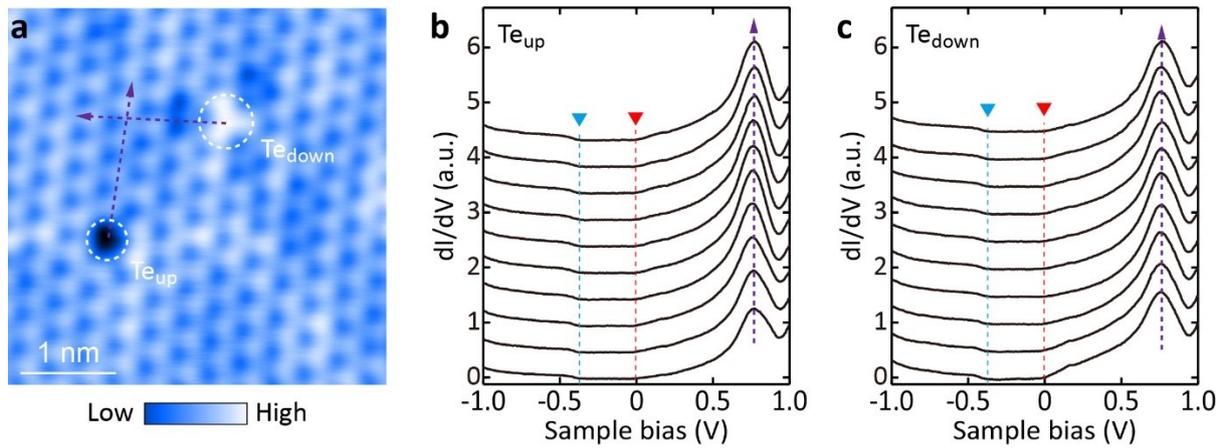

**Figure S2.** Native Te defects in ZrTe$_2$. (a) STM topography ($V$ = 0.1 V, $I$ = 50 pA) of 1 TL ZrTe$_2$ presenting two Te-site defects, Te$_{up}$ and Te$_{down}$, marked by the white circles. (b,c) dI/dV spectra acquired along the two purple dashed lines in (a). The purple arrows mark a prominent DOS peaks at ∼ 0.76 eV, while the red and light blue dashes/triangles denote the CBM and VBM, respectively. Setpoint: $V$ = 1.0 V, $I$ = 200 pA, $\Delta V$ = 30 mV. For clarify the curves have been vertically offset.

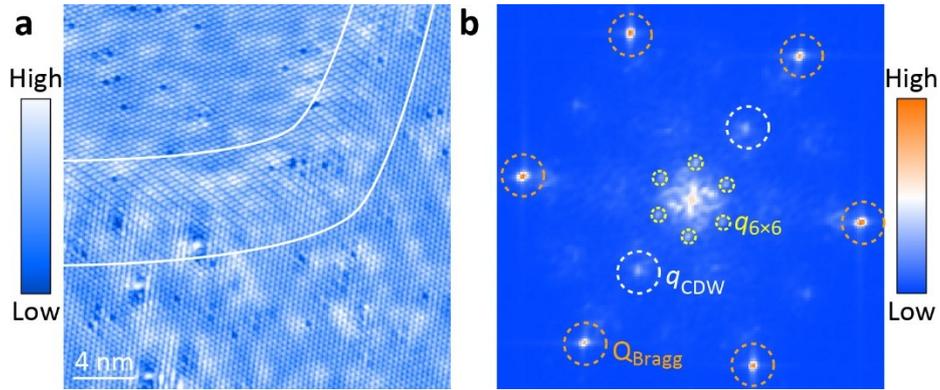

**Figure S3.** Unidirectional CDW modulation in 1 TL ZrTe$_2$. (a) STM topography ($V$ = -10 mV, $I$ = 100 pA) of 1 TL ZrTe$_2$ showing a unidirectional CDW order in the fan shaped region, embraced by the two white lines. (b) FFT image of (a). Q$_{Bragg}$, $q_{CDW}$ and $q_{6\times6}$ correspond to the atomic peak (orange-circled), $2 \times 2$ CDW wave vector (white-circled) and $6 \times 6$ reconstruction of the underlying substrate (yellow-circled), respectively. Note that $q_{CDW}$ occurs only along one of the symmetry equivalent $\Gamma$M directions.

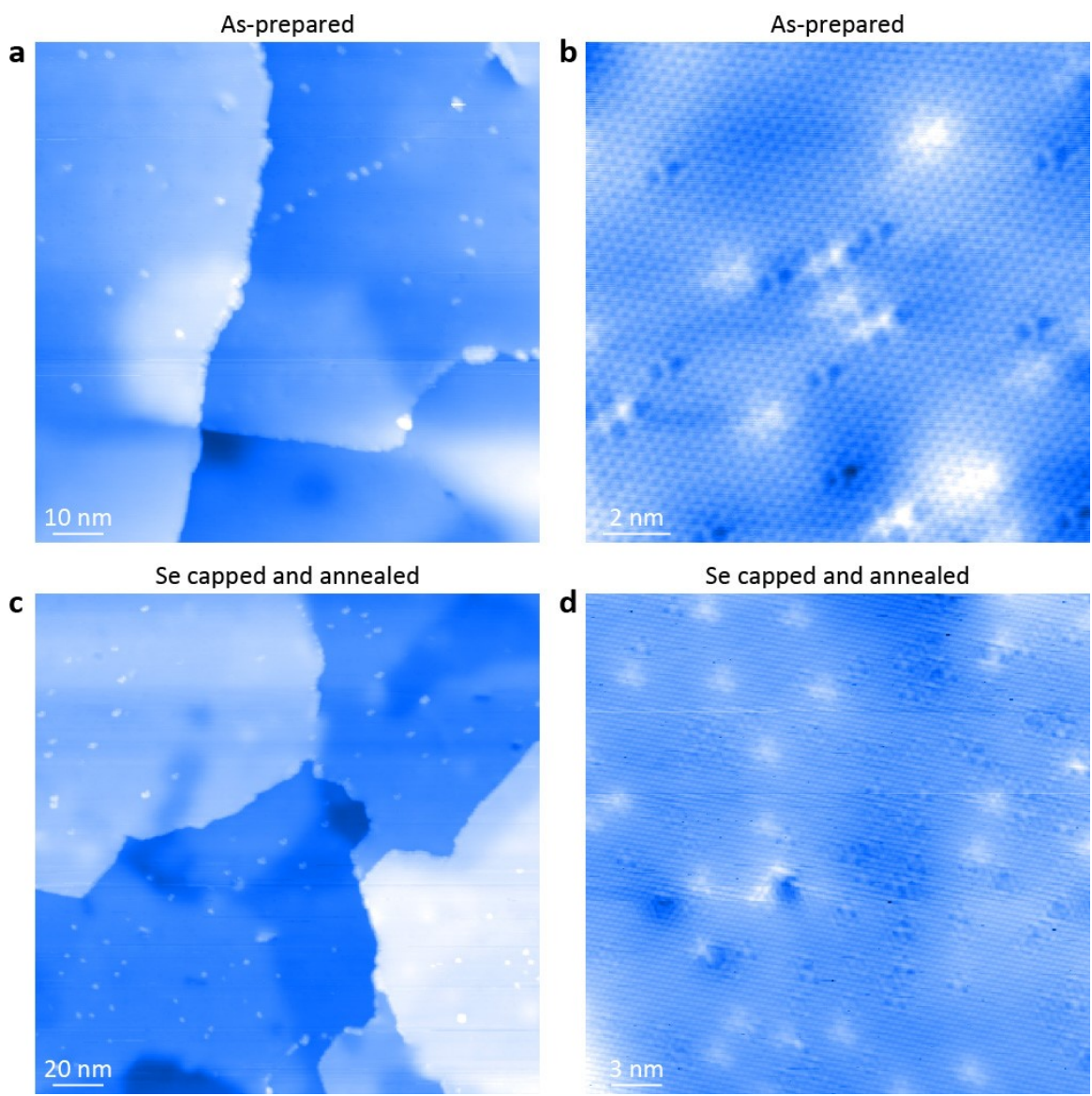

**Figure S4.** Decapping Se protection layers for *ex-situ* ARPES measurements. STM topographic images of (a, b) as-prepared and (c, d) Se-capped and annealed $ZrSe_2$ films with a thickness of 10 TL. The Se-capped and annealed $ZrSe_2$ films show no change with the as-prepared ones, including the defect type and areal density.